# Spin-polarized transport in magnetic tunnel junctions with ZnTe barriers

W. G. Wang [1], C. Ni [2], A. Ozbay [1],  L. R. Shah [1], X. Fan [1],

X.M. Kou [1],  E.R. Nowak [1], and J. Q. Xiao [1]

1)   *Department of Physics and Astronomy, University of Delaware, Newark, DE  19716,USA*

2)   *Department of Materials Science and Engineering, University of Delaware, Newark, DE 19716,USA*

Magnetic tunnel junctions with wide band gap semiconductor ZnTe barrier were fabricated. A very low barrier height and sizable magnetoresistance were observed in the Fe/ZnTe/Fe junctions at room temperature. The nonlinear I-V characteristic curve confirmed the observed magnetoresistance is due to spin-dependent tunneling effect. Temperature dependent study indicated that the total conductance of the junction is dominated by direct tunneling, with only a small portion from the hopping conduction through the defect states inside the barrier.

Magnetic tunneling junctions (MTJs) have wide application in hard disk read head, magnetic random access memory, ultra-low field sensor and many other devices.[1-3] Giant tunneling magnetoresistance (TMR) can be achieved in the MTJs with crystalline MgO barrier.[4,5] In the coherent tunneling process, combination of the symmetry filtering property of MgO and the highly spin-polarized electrons with $\Delta_1$ symmetry in Fe or CoFe electrodes gives rise to the large TMR effect.[6,7] Here most important feature of MgO-(001) tunnel barrier is that the $\Delta_1$ evanescent state has the lowest decay rate, while other evanescent states with $\Delta_5$, $\Delta_2$ or $\Delta_{2'}$ symmetry decay much faster. Compared to traditional $Al_3O_2$ barrier, MgO shows superior performance at small thickness. Therefore MgO barrier is commonly used in the read head of hard disk drives. Though generally speaking for MRAM applications MTJs with perpendicular magnetic anisotropy are preferred,[8-11] for hard disk read heads the focus is still at the resistance-area (RA) product. For example in hard disks with density higher than 2Tb/in², which requires a RA product smaller than $0.3\Omega\mu m^2$ simultaneously with a TMR larger than 50%, the use of MgO tunnel barrier may not be desirable.[12,13] This is primarily due to the large forbidden gap of MgO (7.8 eV),[14] which posts intrinsic limitation for reducing the barrier height below 0.35 eV.[5,15] Therefore, it is of great importance to explore new barrier materials that possess similar electronic structure as MgO but with lower barrier heights.

In this work, we report the study on the room temperature tunneling magnetoresistance in MTJs with ZnTe barriers. Similar to MgO, the lowest orbital of the conduction band in the (001) direction of ZnTe has $\Delta_1$ symmetry. A large TMR was predicated with ZnSe,[16] which shares very similar electronic band structure with ZnTe. Compared common barrier materials such as $Al_3O_2$ and MgO, ZnTe has a much lower band gap of 2.2 eV.[17] Therefore, a large TMR and small RA could potentially be realized in junctions with ZnTe barriers. ZnTe has a stable zinc blende phase with a lattice constant of 0.61nm.[17] It has important applications in Terahertz imaging,[18] tunneling diode,[19] solar-cells contacts[20] and diluted magnetic semiconductors.[21,22] Our study shows that a very low barrier height (0.13 eV) can be obtained in junctions with ZnTe barriers. A tunneling magnetoresistance of 3.5% is observed at RT. The spin-polarized transport across the Fe/ZnTe interface demonstrated here also paves the way for studying spin manipulation in ZnTe at room temperature.

Samples in this study were fabricated in a vacuum chamber equipped with 6 sputtering sources, one plasma source and one ion source. The base pressure of the system is 5 x 10$^{-8}$ Torr. The sample structures were Si/Cu 70/ IrMn 15 / CoFe 6/Fe 1.6/ZnTe(1-3)/Fe 1.6/CoFe 12/Cu 50 (numbers indicate layer thickness in nanometers). The Si(001) wafers were treated by HF and properly cleaned before loading into the chamber. The layers below ZnTe barrier were chosen so that the bottom Fe interface is (001) oriented. In the current study, the barrier layers were deposited at RT and 160°C. After the fabrication of blanket films, MTJs with diameters ranging from 5μm to 100μm were defined by standard microfabrication process including multiple steps of photolithography and ion-beam milling. The RT transport measurement was performed in the standard 4-probe configuration on a probe station. The low temperature properties were tested in a PPMS system. More details in the MTJ fabrication and testing techniques are described by our previous publications.[23-25]

The structure of unpatterned MTJ films was first characterized by x-ray diffraction. As shown in Figure 1(a), the film stack below ZnTe barrier is (001)-oriented. The CoFe(001) layer was grown on IrMn(001) despite a large lattice mismatch of 6.3%.[26] The Cu (311) peak at about 90 degree is from the capping layer as it is absent in the sample without top Cu layers ( not shown).  The lattice mismatch between Fe(001) and ZnTe(001) is 6.4%. Therefore a Fe(001)/ZnTe(001)/Fe(001) epitaxial sandwich structure could possibly be achieved just as the case of Fe(001)/MgO(001)/Fe(001) junctions, thus lead to high TMR ratios.[6,16] The cross-sectional transmission electron microscopy (TEM) study was carried out to investigate the microstructure of the barrier layer. A uniform ZnTe barrier with very small roughness can be grown on the Fe electrode as shown in Figure 1(b). The distinct interfaces between the ZnTe barrier and the Fe electrodes can be clearly seen. However, the ZnTe barrier exhibits amorphous nature as shown in the high resolution image in Figure 1(c). The barrier layer was also found amorphous when ZnTe was deposited at 160°C. The RF sputtering power and Ar pressure used for the ZnTe barrier was 40 W and 4.5 mTorr, respectively. Further investigation on the effect of sputtering power, inner gas pressure and substrate temperature is necessary to optimize the barrier structure.

The room temperature TMR ratios for junctions with difference barrier thickness are shown in Figure 2(a). TMR values around 2% were obtained in a wide range of barrier thicknessses when ZnTe was deposited at RT. On the other hand, TMR quick decreases with reducing barrier thickness when ZnTe was deposited at 160°C. The optimal barrier thickness for the highest TMR is 2.3nm and 2.6nm, respectively, for barriers deposited at RT and 160°C. The dependence of the antiparallel RA of the MTJs on the barrier thickness is shown in Figure 2(b). Due to the much lower band gap of ZnTe, the RAs in ZnTe MTJs are generally more than 3 orders of magnitude lower than the RAs in MgO junctions with the same barrier width. The tunneling resistance in junctions with RT-deposited barrier scales almost linearly with barrier thickness in the log scale. In the WKB approximation, the slope of resistance versus barrier thickness curve can be expressed as $4\pi(2m\phi)^{1/2}/h$,[27] where m is the electron mass, $\phi$ is the barrier height and $h$ is the Planck's constant. From the slope we can estimate the barrier height of ZnTe is about 0.13 eV. This value is substantially lower than MgO (0.35-0.4 eV) [5,15] and much lower than ZnS (~0.58 eV).[28] The TMR ratios can be improved by post-growth thermal annealing. The highest TMR in the present study is 3.5% after the MTJs were annealed at 200°C as shown in Figure 2(c). The TMR curve show a very sharp switch at around 25 Oe, indicating very flat barrier-electrode interfaces as observed in the TEM pictures. The tunneling characteristics is confirmed by the nonlinear IV curve of the junction as shown in the inset of Figure 2(c), which demonstrates the observed magnetoresistance is due to tunneling instead of GMR effect.

It is important to characterize the percentage of spin independent contribution to the total conductance due to thermal excitation in our junctions with semiconducting barriers. The temperature dependence of TMR is shown in Figure 3(a). The TMR almost linearly increases with decreasing temperature. In general, the total conductance in a MTJ can be expressed by the model proposed by Shang *et al* as, [29]

$$G(\theta) = G_T(1 + P_1 P_2 \cos\theta) + G_{SI} \qquad\qquad Eq.\ (1)$$

, where the first term is form the direct spin-dependent tunneling; the second term, $G_{SI}$, is from spin-independent conduction. The direct tunneling conductance coefficient $G_T$ only depends weakly on temperature and can be treated as constant. [29] Then for our junction with Fe as both top and bottom electrodes the TMR can be described by the extended Julliere's formula,

$$TMR = 2P^2/(1 - P^2 + G_{SI}/G_T). \qquad\qquad Eq.\ (2)$$

By assuming the spin polarization of the ferromagnetic electrode has the same temperature dependence as surface magnetization(thus following the $T^{3/2}$ rule), the conductance difference between parallel and antiparallel configuration can be expressed as $\Delta G = 2G_T\ P^2 = 2G_T\ P_0^2\ (1 - \alpha\ T^{3/2})^2$, where $P_0$ is the zero temperature spin polarization at the Fe/ZnTe interfaces. $\Delta G$ can be fit with the equation above as shown in Figure 3(b). The best fit value for $\alpha$ is 2 x $10^{-5}$ $K^{-3/2}$, consistent with previous report.[30] Now the spin independent conductance due to thermal excitation can be calculated by $G_{SI}$ = ⟨ $G$ ⟩ - $G_T$ , where ⟨ $G$⟩ is the average conductance between parallel and antiparallel configuration. $G_T$ can be derived from the fitting of $\Delta G$. The temperature dependence of $G_{SI}$ is shown in the inset of Figure 3(b). The behavior of $G_{SI}$ can be explained by the electron hopping through the defect states inside the ZnTe barrier. The hopping process is greatly activated upon increasing temperature. $G_{SI}$ can be well fit by $\sigma_0 + \sigma_2 T^{1.33}$, corresponding to the hopping through 2 localized states in the ZnTe barrier.[31] The defect states assisted hopping conduction was also observed in junctions with $Al_2O_3$,[23,29] MgO [30] and ZnSe [32] barrier. It also has a large impact to the noise level of the MTJs.[33] In the early study on MTJs with ZnSe barrier, it was found spin independent conduction dominates at RT due to the defect states assisted hopping.[34] In our junctions $G_{SI}$ only accounts for about 10% of total conductance at RT as shown in the inset of Figure 3(b), demonstrating good barrier quality despite the amorphous nature of ZnTe. The TMR in this system now can be readily calculated by using Equation 2. The experimental data can be well represented by the calculation as shown in Figure 3(a).

To summarize, we have investigated the spin-dependent transport properties in MTJs with ZnTe as barriers. ZnTe has been identified as a potential candidate for achieving ultra-low RA values in MTJs. The barrier height of ZnTe estimated by WKB approximation is only 0.13 eV, much lower than most other tunneling barrier. A TMR ratio of 3.5% was observed at RT. Upon further optimization, a substantial enhancement of TMR is possible in junctions with crystalline ZnTe barrier through the coherent tunneling mechanism. The spin-polarized transport across the Fe/ZnTe interface demonstrated here will also help future study on manipulation of spin in ZnTe.

## ACKNOWLEDGEMENT

This work was supported by DOE DE-FG02-07ER46374 and NSF Grant No. DMR0827249.

FIGURES CAPTIONS

Figure 1(color online). (a). θ-2θ X-ray diffraction pattern of the MTJ film. (b) Cross-sectional TEM image of the sample. (c) High resolution TEM image of the sample for the barrier region.

Figure 2(color online). (a) Dependence of TMR on the barrier thickness for MTJs, with ZnTe barrier fabricated at RT (square) and 160°C (dot). (b) Dependence of RA on the barrier thickness. The straight lines are the fit by the WKB approximation for determining the barrier height. (c) TMR curve of the MTJ after post-growth annealing at 200°C. Inset shows the I-V curve of the junction.

Figure 3(color online). (a) Temperature dependence of TMR (square) and the fit by the extended Julliere formula (solid line). (b) Temperature dependence of $\Delta G$ and the fit based on $T^{3/2}$ rule. Inset shows the ratio of the spin independent conductance to the total conductance and the fit based on the defect states assisted hopping picture.

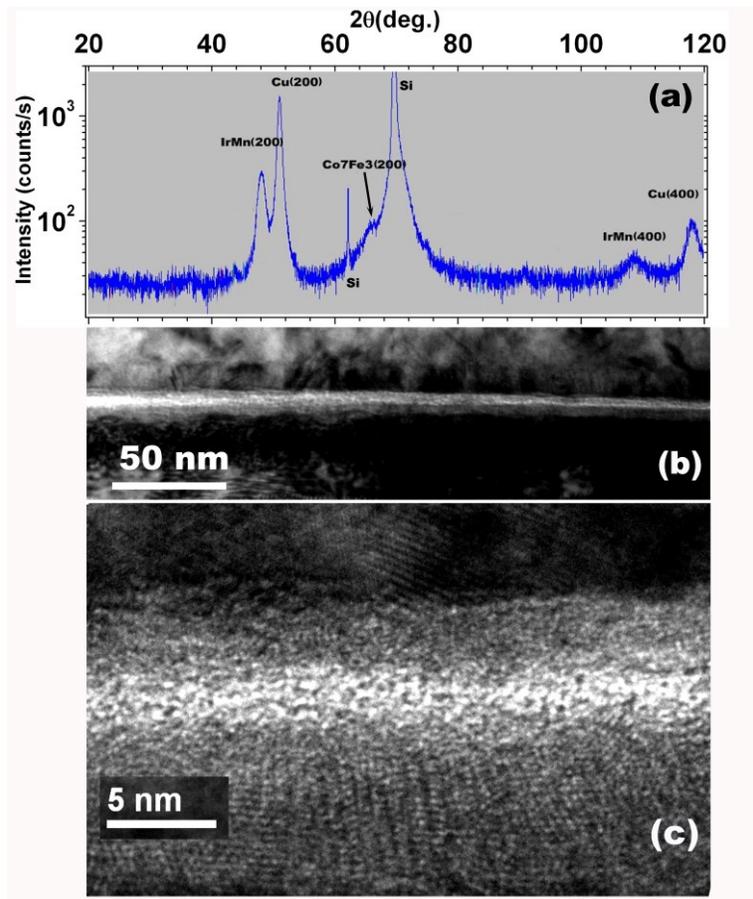

Figure 2(color online). (a). θ-2θ X-ray diffraction pattern of the MTJ film. (b) Cross-sectional TEM image of the sample. (c) High resolution TEM image of the sample for the barrier region.

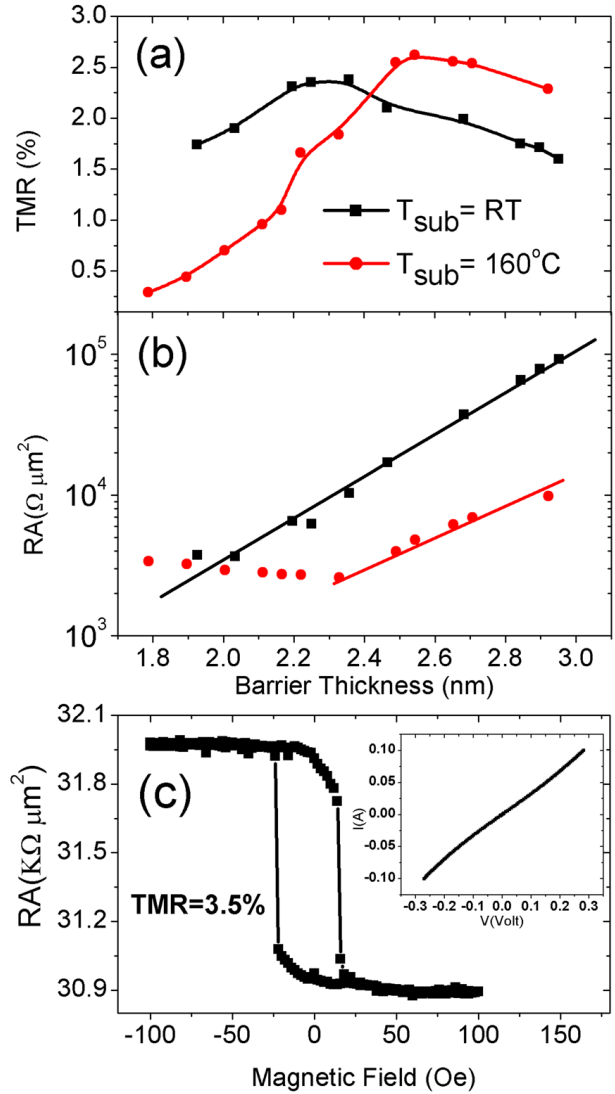

Figure 2(color online). (a) Dependence of TMR on the barrier thickness for MTJs, with ZnTe barrier fabricated at RT (square) and 160°C (dot). (b) Dependence of RA on the barrier thickness. The straight lines are the fit by the WKB approximation for determining the barrier height. (c) TMR curve of the MTJ after post-growth annealing at 200°C. Inset shows the I-V curve of the junction.

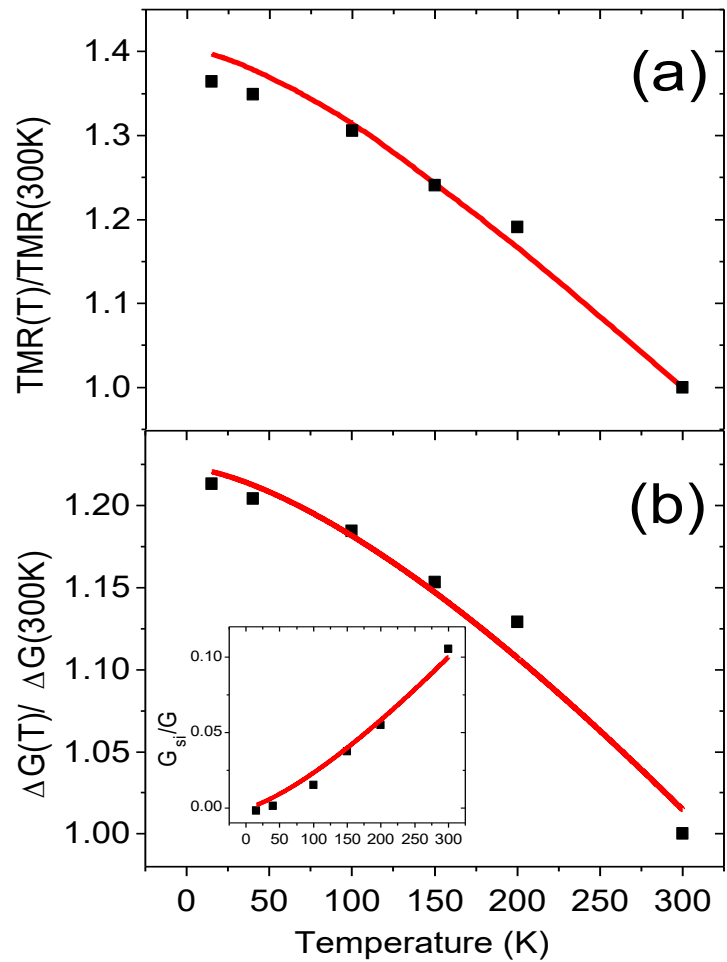

Figure 3 (color online). (a) Temperature dependence of TMR (square) and the fit by the extended Julliere formula (solid line). (b) Temperature dependence of ∆G and the fit based on T$^{3/2}$ rule. Inset shows the ratio of the spin independent conductance to the total conductance and the fit based on the defect states assisted hopping picture.